\documentclass[reprint,aps,prd,floatfix,nofootinbib,superscriptaddress,showpacs]{revtex4-2}
\usepackage{amsmath,amssymb}
\usepackage{graphicx}
\usepackage{booktabs}
\usepackage{hyperref}
\usepackage[left,mathlines]{lineno}
\usepackage[none]{hyphenat}
\hypersetup{
    hidelinks
}

\begin{document}
\title{The Generalization Gap in Machine Learning EoS Inference from Core-Collapse Supernova Gravitational Waves}

\author{Ayan Mitra}
\email{ayan@illinois.edu}
\affiliation{Center for AstroPhysical Surveys, National Center for Supercomputing Applications, Urbana, IL 61801, USA}
\affiliation{Department of Astronomy, University of Illinois at Urbana Champaign, Urbana, IL 61801, USA}

\date{\today}

\begin{abstract}
Core collapse supernova gravitational waves may carry information about the dense matter equation of state (EoS), where dense matter means nuclear matter compressed to roughly atomic nucleus density or above, and the EoS describes how pressure, density, temperature, and composition are related. This makes the scientific question a test of physical inference: can a machine learning model trained on a finite waveform catalogue, a collection of simulated gravitational wave time series, predict EoS parameters for an EoS family that was absent from training? Under standard random cross validation, where the catalogue is randomly split into training and test sets, a LightGBM gradient boosted decision tree regressor appears successful, giving the coefficient of determination $R^2=(0.70,0.67,0.60)$ for $(K_0,J,L)$, the nuclear incompressibility, symmetry energy, and slope of the symmetry energy. However, under Leave One EoS Out validation, where all waveforms from one EoS are withheld from training, the same model gives mean absolute errors of $(44.57,3.19,30.54)$  and negative pooled $R^2$ values, meaning worse than predicting the mean target value. The failure persists across linear models, random forests, neural networks, and gradient boosted trees. Restricting the input to low dimensional physical features such as bounce amplitude, bounce width, and post bounce peak frequency reduces template leakage, the memorisation of closely related templates shared between training and test sets, but does not restore reliable out of catalogue EoS inference. A progenitor mass case study shows a different behaviour: classification generalises to unseen rotation speeds, while continuous mass regression compresses predictions towards the catalogue interior. These results show that machine learning can interpolate within current CCSN waveform catalogues, but this does not imply robust physical inference for unseen EoS models. Future CCSN machine learning pipelines should therefore use leave family out validation, richer simulation coverage, and physics aware inference frameworks.
\end{abstract}

\maketitle

\section{Introduction}
Core collapse supernovae (CCSNe) represent some of the most energetic stellar phenomena in the universe, marking the evolutionary end of massive stars ($M \gtrsim 8 M_\odot$). During the collapse, the inner core reaches densities exceeding nuclear saturation, the characteristic density of atomic nuclei ($\rho_0 \approx 2.7 \times 10^{14}\text{ g/cm}^3$). This is the dense matter regime: matter is compressed so strongly that nuclear physics, rather than ordinary gas physics, determines how it responds. Core collapse therefore provides a unique laboratory for studying the dense matter equation of state (EoS; also written EoS) under conditions inaccessible to terrestrial laboratories. The EoS specifies how pressure, density, temperature, and composition are related in dense nuclear matter, and therefore controls how the proto neutron star core compresses and oscillates~\cite{oertel2017}.

The gravitational wave (GW) signature emitted during the core bounce and early post bounce phase carries information about the newborn proto neutron star (PNS), rotation, progenitor structure, and the dense matter EoS~\cite{richers2017, abdikamalov2014, oconnor2018,vartanyan2023,choi2024}. In this paper a waveform means the time series of simulated GW strain around the bounce and early post bounce signal. Richers et al.~\cite{richers2017} showed, however, that the bounce signal itself is largely independent of the EoS and is primarily sensitive to $T/|W|$, the ratio of rotational kinetic energy to gravitational binding energy, and to differential rotation, where different parts of the core rotate at different rates. Stronger EoS dependence appears in the post bounce oscillation frequency $f_{\rm peak}\sim 600$ to $1000$ Hz. Modern long term 3D simulations further show that PNS oscillation modes, accretion dynamics, angular anisotropy, matter memory, and neutrino memory all contribute to the observable CCSN GW phenomenology~\cite{vartanyan2023,choi2024,burrows2024}. This already suggests that EoS inference from CCSN GWs is a degenerate inverse problem rather than a direct waveform to label lookup.

With the upcoming development of third generation (3G) interferometers, such as the Einstein Telescope (ET)~\cite{et_cite} and Cosmic Explorer (CE)~\cite{ce_cite,reitze2019}, we expect greatly improved access to Galactic and nearby CCSNe. Several studies quantify this opportunity. Srivastava et al.~\cite{srivastava2019} found that even a supernova optimised CE like detector is limited to roughly $\mathcal{O}(100)$ kpc for likely CCSN waveforms, corresponding to a Galactic scale event rate of about two per century, and that a one per year CCSN GW rate would require strain sensitivity near $3\times 10^{-27}\,{\rm Hz}^{-1/2}$ over $100$ to $1500$ Hz. Afle and Brown~\cite{afle2020} found that, for a rapidly rotating Galactic centre event at $8.1$ kpc, CE could constrain the rotation parameter $\beta$ to about $8\times 10^{-4}$ and the post bounce oscillation frequency to within $5$ Hz at 90\% credibility; at the Magellanic Cloud distance of $48.5$ kpc these degrade to about $0.003$ in $\beta$ and $11$ Hz. These are precisely the regimes in which ML pipelines will be tempting, but also where validation failures would be scientifically costly.

In recent years, machine learning (ML) has been widely proposed for inferring dense matter parameters directly from CCSN GW waveforms~\cite{mitra2022, mitra2024, abylkairov2024, nunes2024}. Many of these works report classification and regression accuracies exceeding 90\%. Here we do not mean that there is a single formal ``ML for CCSN programme.'' Rather, we use this phrase to describe a sequence of related recent studies that progressively add more realistic assumptions to CCSN GW machine learning: detector noise, larger progenitor coverage, uncertain bounce time, and parameter estimation. Abylkairov et al. (2025)~\cite{abylkairov2025} investigated the maximum distance at which various detectors, including Advanced LIGO A+ (aLIGO A+), ET, and CE, could classify equations of state. Wang et al. (2026)~\cite{wang2026} proposed supervised contrastive learning, a neural network training strategy that pulls examples from the same class closer together in latent space, with ResNet 50 architectures to distinguish CCSN signals from detector noise. More recently, Akhmetali et al. (2026a)~\cite{akhmetali2026a} relaxed several key assumptions by training EoS classifiers on multi progenitor catalogues under core bounce time uncertainties. In a subsequent study, Akhmetali et al. (2026b)~\cite{akhmetali2026b} performed parameter estimation (peak frequency, peak amplitude, core rotation) and analysed detector horizons.

While these works show promising results for classification and core parameter estimation, they typically evaluate their models using standard random cross validation (CV). Cross validation means splitting the catalogue into training and testing subsets, repeating the process across several folds, and averaging the test performance. In catalogues where many waveforms share the same EoS or progenitor model but differ in rotation profile, random splits can place closely related templates in both training and test sets. This allows the model to memorise the geometric manifolds of specific templates (known as template leakage) rather than learning transferable physical correlations. This is the CCSN waveform version of a broader leakage problem in machine learning: examples that are not truly independent can appear in both the training and test sets, causing overly optimistic performance. Similar failures have been documented across machine learning based science~\cite{kapoor2023} and in medical imaging, where nearby image slices or images from the same subject can leak between splits~\cite{tampu2022}. These communities mitigate the problem by using subject wise, group wise, or time aware validation splits instead of random splits when samples share a hidden parent source.

Previous studies established that simulated CCSN GW catalogues contain learnable structure~\cite{engels2014,edwards2021,chao2022}. Our goal is simple: we ask whether a model has learned a physical rule that could work for a new EoS, or whether it has mostly learned to recognise patterns already present in the finite simulation catalogue. The present work asks a stricter question: whether the learned mappings remain valid for EoS models and rotation configurations absent from the training catalogue. In this paper, \textit{catalogue interpolation} denotes random training and test splits within the same simulation families, \textit{physically grouped validation} denotes splits that hold out entire physical groups, and \textit{out of distribution} performance denotes prediction on those held out groups. The title phrase ``out of catalogue'' refers to this finite catalogue setting, where the held out physical group is absent from the training simulations. Leave One EoS Out (LOEO) validation is the EoS specific grouped validation used for the main dense matter regression test: each fold trains on all but one EoS and tests on the omitted EoS. The mission of this paper is therefore to turn a technical validation issue into a physical question: can the model make a reliable prediction for an EoS family it has never seen before? This work should therefore be viewed as a generalisation audit of recent CCSN GW machine learning studies: rather than asking whether ML models can interpolate within an existing catalogue, we ask whether they extrapolate to unseen EoS models and unseen rotation configurations. We analyse the generalisation gap between random cross validation and out of distribution EoS validation across multiple machine learning architectures. We show that extracting physical observables narrows this gap in some cases, and contrast it with progenitor mass classification, which generalises robustly to unseen rotation profiles.

\section{Waveform Catalogues and Targets}
We analyse two public benchmark catalogues that have been used in recent CCSN GW machine learning studies:
\begin{enumerate}
    \item \textbf{Richers et al. (2017) Catalogue~\cite{richers2017}}: This catalogue contains strains from 1,824 axisymmetric simulations spanning 98 rotation profiles ($\omega_0 \in [0, 15.5]\text{ rad/s}$ and $A \in [300, 10000]\text{ km}$), using a single $12 M_\odot$ progenitor. The data file used here is the public Zenodo HDF5 release \texttt{GWdatabase.h5}~\cite{richersZenodo}. The original study describes 18 EoS models; the HDF5 labels used here contain 21 EoS strings because several electron capture and related variants appear as distinct labels. These waveforms were generated with CoCoNuT~\cite{dimmelmeier2002,dimmelmeier2005}, a general relativistic hydrodynamics code for rotating stellar collapse that uses the conformal flatness condition (CFC) for the spacetime metric and high resolution shock capturing hydrodynamics in spherical polar coordinates. Richers et al. used CoCoNuT in axisymmetry with parameterised deleptonisation during collapse, approximate neutrino leakage after bounce, and quadrupole GW extraction~\cite{richers2017,abdikamalov2014}. Thus the dataset should be read as a 2D CFC general relativistic CoCoNuT waveform catalogue rather than a full 3D neutrino radiation hydrodynamics population.
    \item \textbf{Mitra et al. (2022) Catalogue~\cite{mitra2022}}: 402 waveforms spanning 4 progenitor masses ($12, 15, 27, 40 M_\odot$) under varying rotation speeds. The data file used here is the public Zenodo CSV release \texttt{GW\_data.csv}~\cite{mitraZenodo}. The Mitra waveforms are also CoCoNuT simulations~\cite{dimmelmeier2002,dimmelmeier2005}, not observational signals. The progenitor structures are based on standard presupernova models~\cite{woosley2002,heger2005,woosley2007}; the simulations use the SFHo nuclear EoS~\cite{steiner2013}, parameterised deleptonisation~\cite{liebendoerfer2005}, and the same CoCoNuT code lineage cited above.
\end{enumerate}

This focus on CoCoNuT should not be read as a claim that all CCSN GW analyses use the same simulation framework. CoCoNuT and its rotating collapse catalogue lineage provide the public benchmark behind much of the recent early bounce and EoS machine learning literature audited here~\cite{richers2017,edwards2021,mitra2022,mitra2024,nunes2024,pastormarcos2024}. The broader CCSN GW literature also uses other waveform families and simulation codes, including neutrino driven multidimensional simulations, long term 3D radiation hydrodynamics models, memory calculations, and detector reconstruction catalogues~\cite{oconnor2018,gossan2016,szczepanczyk2021}~\cite{vartanyan2023,choi2024,burrows2024}. In particular, Burrows and collaborators provide a public high cadence 3D waveform repository containing matter strains, matter quadrupole tensors, neutrino memory strains, and spectra for a suite of long term CCSN models~\cite{burrowsGWData}. We therefore focus on CoCoNuT because it is the common catalogue basis for the EoS and mass studies analysed in this paper; the same grouped validation logic should be applied to other simulation catalogues, with the held out group chosen according to the relevant physical family, such as progenitor, EoS, rotation law, dimensionality, simulation code, viewing angle, or memory component.

To isolate the bounce and early ringdown phases in the Richers EoS catalogue, waveforms are cropped from $2.0$ ms before core bounce to $6.0$ ms after core bounce, which yields 524 samples at $65.5\text{ kHz}$ sampling frequency. After excluding 60 Richers simulations that do not contain this full bounce window, the resulting EoS regression set contains 1,764 standard Richers signals. The Mitra mass catalogue is analysed separately and contains 402 waveforms.

For the Richers EoS regression task, the 21 unique EoS labels are mapped to their continuous bulk nuclear saturation parameters. These parameters summarise the behaviour of cold nuclear matter near saturation density, where many EoS models are calibrated:
\begin{itemize}
    \item \textbf{Incompressibility modulus ($K_0$, in MeV)}: Measures how hard it is to compress symmetric nuclear matter near saturation density; larger values correspond to a stiffer response.
    \item \textbf{Symmetry energy ($J$, in MeV)}: Measures the energy cost of changing symmetric matter into neutron rich matter at saturation density.
    \item \textbf{Symmetry energy slope ($L$, in MeV)}: Measures how rapidly the symmetry energy changes with density near saturation.
\end{itemize}

\begin{figure}[t]
    \centering
    \includegraphics[width=\columnwidth]{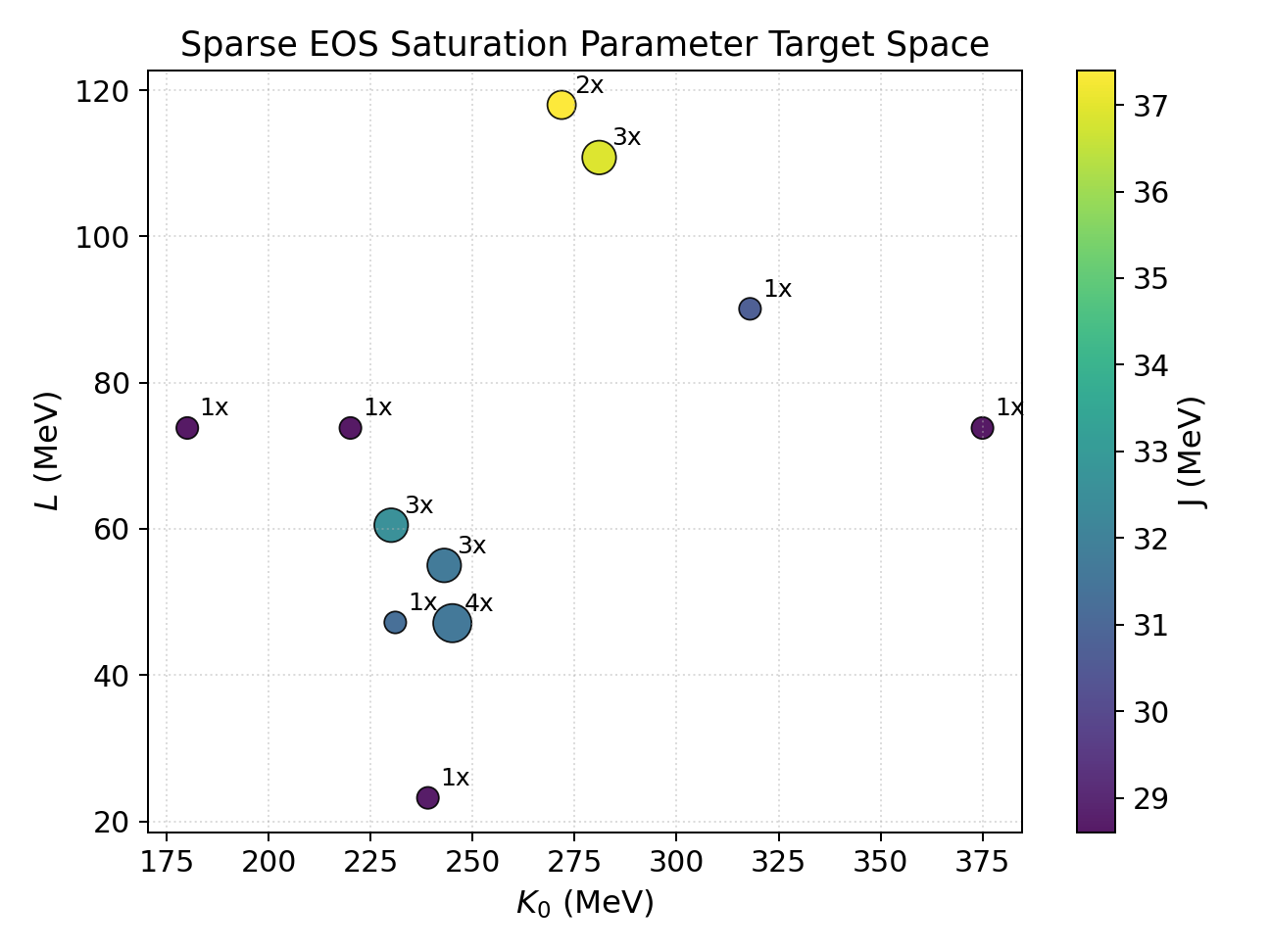}
    \caption{Sparse saturation parameter target space for the EoS labels used in this analysis. Marker colour gives $J$, marker size and labels indicate how many EoS strings share the same $(K_0,J,L)$ triplet.}
    \label{fig:eos_target_space}
\end{figure}

The target space is sparse. The 21 EoS strings map to only 11 unique $(K_0,J,L)$ triplets because several EoS labels share identical saturation parameters. Names such as BHBL, BHBLP, HSDD2, and SFHo are catalogue labels for particular finite temperature nuclear EoS tables, not new observables measured from the waveform. Some labels differ by composition, table construction, or electron capture treatment while sharing the same saturation constants. For example, BHBL, BHBLP, and HSDD2 all map to $(243.0,31.7,55.0)$ MeV, while SFHo and three SFHo electron capture variants all map to $(245.0,31.6,47.1)$ MeV. Figure~\ref{fig:eos_target_space} shows this target grid. This sparsity is a limitation for continuous regression, but it also makes the failure under LOEO more striking: in several held out folds, the training set still contains EoS labels with the same target triplet, yet the model cannot reliably transfer the waveform to parameter mapping.

\section{The EoS Generalisation Gap}
To systematically audit the generalisation capabilities of machine learning models, we evaluate multiple algorithms representing different model families:
\begin{enumerate}
    \item \textbf{Ridge Regression}: A linear regression model with an $L_2$ penalty that discourages overly large fitted weights~\cite{hoerl1970}.
    \item \textbf{Random Forest (RF)}: An ensemble method that trains many decision trees on resampled data and averages their predictions~\cite{breiman2001}.
    \item \textbf{Multi Layer Perceptron (MLP)}: A feed forward artificial neural network made of stacked nonlinear layers~\cite{rumelhart1986}.
    \item \textbf{LightGBM}: Light Gradient Boosting Machine, a fast gradient boosted decision tree architecture that builds an ensemble of decision trees to improve regression or classification accuracy~\cite{ke2017}.
\end{enumerate}

We evaluate these models across two validation protocols:
\begin{enumerate}
    \item \textbf{Random 5 Fold CV}: Standard cross validation where waveforms are shuffled randomly. We report the mean $R^2$ score and its standard deviation across folds to quantify fold to fold scatter.
    \item \textbf{Leave One EoS Out (LOEO) CV}: The model is trained on 20 EoS labels and evaluated on the single remaining, entirely unseen EoS label. This is repeated 21 times to compute the overall out of distribution generalisation metrics.
\end{enumerate}

A critical concern is the size of the target label space. The Richers catalogue provides 1,764 usable waveforms for this analysis, but these waveforms occupy only 21 EoS labels and only 11 unique $(K_0,J,L)$ target triplets. The effective regression grid is therefore 11 points in a three dimensional target space, with many waveforms repeated at each point while rotation changes. This is sparse for continuous regression because the model must learn how the waveform changes across $K_0$, $J$, and $L$, while the training data sample only a few target locations. In each LOEO fold, all test instances share the same ground truth nuclear parameters, so a fold wise $R^2$ is undefined and a pooled $R^2$ can be sensitive to the catalogue target distribution. By pooled $R^2$ we mean a single coefficient of determination computed after concatenating the predictions from all LOEO folds and comparing them with the corresponding true values. This gives one summary number for the whole held out EoS set, even when the per fold score is not informative. We therefore use MAE, mean predictor baselines, and per EoS error diagnostics as the primary evidence, while still reporting pooled $R^2$ to make the failure directly comparable with random CV scores.

Here $R^2$ denotes the coefficient of determination: $R^2=1$ is a perfect prediction, $R^2=0$ is equivalent to predicting the global mean target value, and $R^2<0$ means the model is worse than that mean predictor. MAE denotes mean absolute error, the average absolute difference between predicted and true values, reported in MeV for the EoS parameters.

\begin{figure}[t]
    \centering
    \includegraphics[width=\columnwidth]{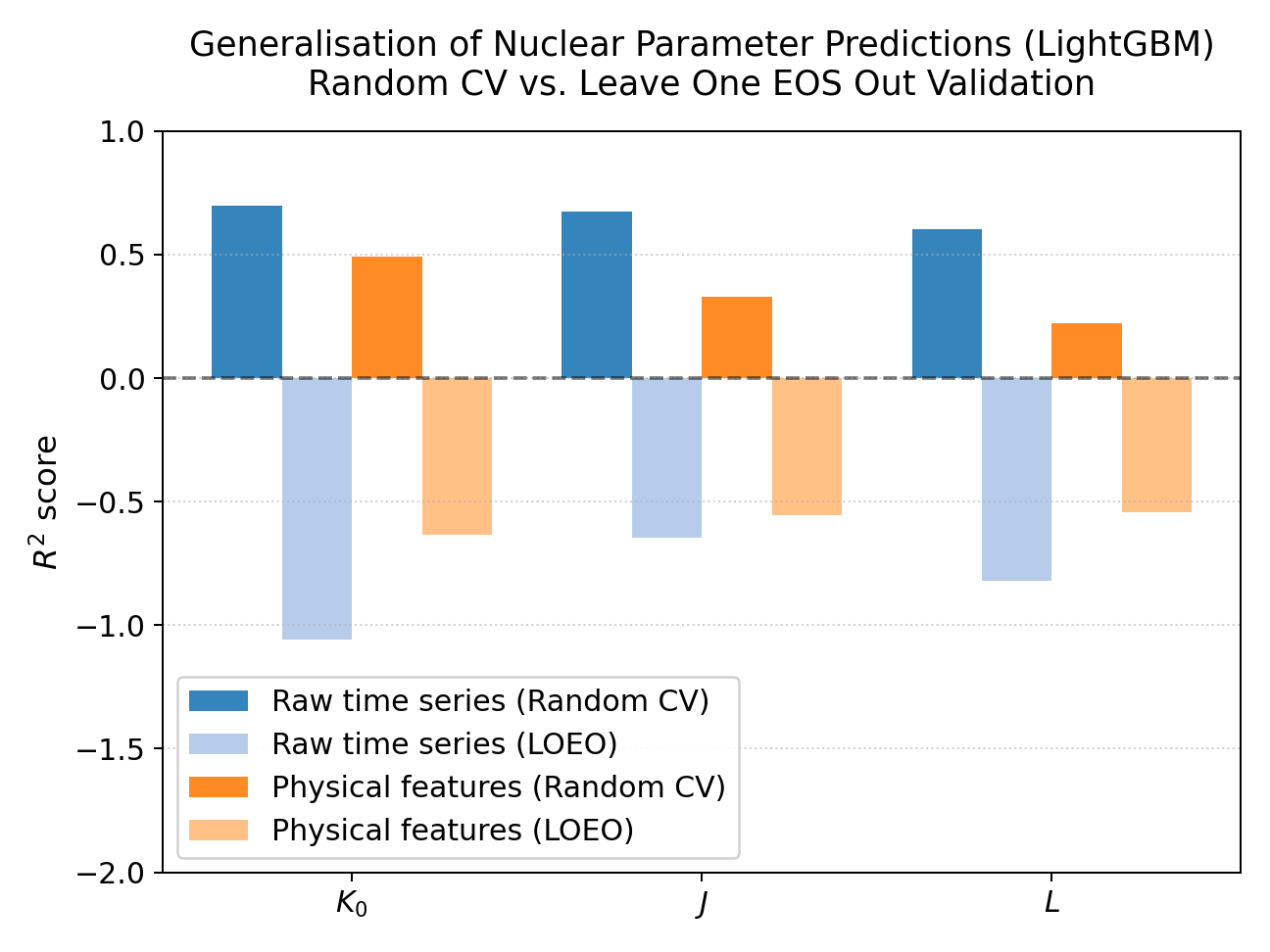}
    \caption{$R^2$ score comparison between random cross validation and Leave One EoS Out validation for $K_0, J,$ and $L$. The raw time series results use LightGBM directly on the 524 sample bounce window; the physical feature results use bounce amplitude, bounce width, and peak post bounce frequency. Scoring below 0 indicates performance worse than predicting the global target mean.}
    \label{fig:gap_eos}
\end{figure}

As shown in Fig.~\ref{fig:gap_eos}, the performance of the LightGBM regressor drops dramatically when moving from Random CV to out of distribution validation. On raw time series, random 5 fold CV gives $R^2=(0.698,0.674,0.601)$ for $(K_0,J,L)$, whereas explicit LOEO validation gives MAE $(44.57,3.19,30.54)$ MeV and pooled $R^2=(-1.059,-0.647,-0.821)$. A global mean predictor has $R^2=0$ by construction and MAE $(31.19,2.47,23.35)$ MeV; a LOEO training mean predictor gives $R^2\simeq -0.12$ for all three targets. Thus the learned raw waveform model performs substantially worse than simple mean baselines when evaluated on unseen EoS labels. This indicates that the model is not learning a transferable nuclear physics mapping, but is instead exploiting template level structure that is shared across random training and test splits.

\begin{figure*}[t]
    \centering
    \includegraphics[width=0.98\textwidth]{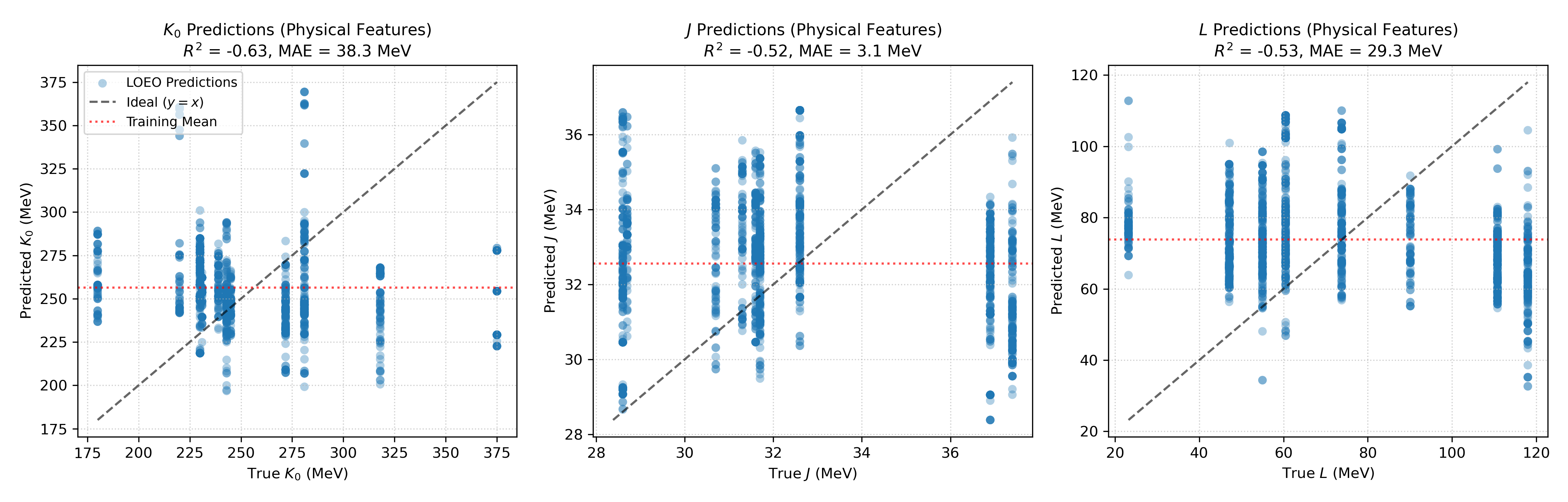}
    \caption{Predictions outside the training sample vs. true values for $K_0$, $J$, and $L$ under Leave One EoS Out (LOEO) cross validation using the LightGBM model trained only on the physical observables. The black reference line represents the ideal $y=x$ case, and the red reference line indicates the mean target value of the training set. The collapse of the predictions towards the training mean is reflected in negative pooled $R^2$ values and reveals the failure of the model to learn a generalisable physical mapping.}
    \label{fig:loeo_predictions}
\end{figure*}

A target distribution diagnostic shows how much of this failure is caused by boundary extrapolation. For $K_0$, the catalogue boundaries are defined by LS180 ($K_0 = 180$ MeV) and LS375 ($K_0 = 375$ MeV). When either is held out, the model must extrapolate and the mean per EoS MAE rises to $120.86$ MeV. For $L$, the held out boundary case SFHx ($L=23.2$ MeV) gives a mean per EoS MAE of $61.99$ MeV. These boundary failures are expected in a sparse catalogue and should not be overinterpreted as a fundamental impossibility result. The collapse towards the mean is visually demonstrated in Fig.~\ref{fig:loeo_predictions}, which shows the predictions under LOEO validation vs. the true target values.

\begin{figure}[t]
    \centering
    \includegraphics[width=\columnwidth]{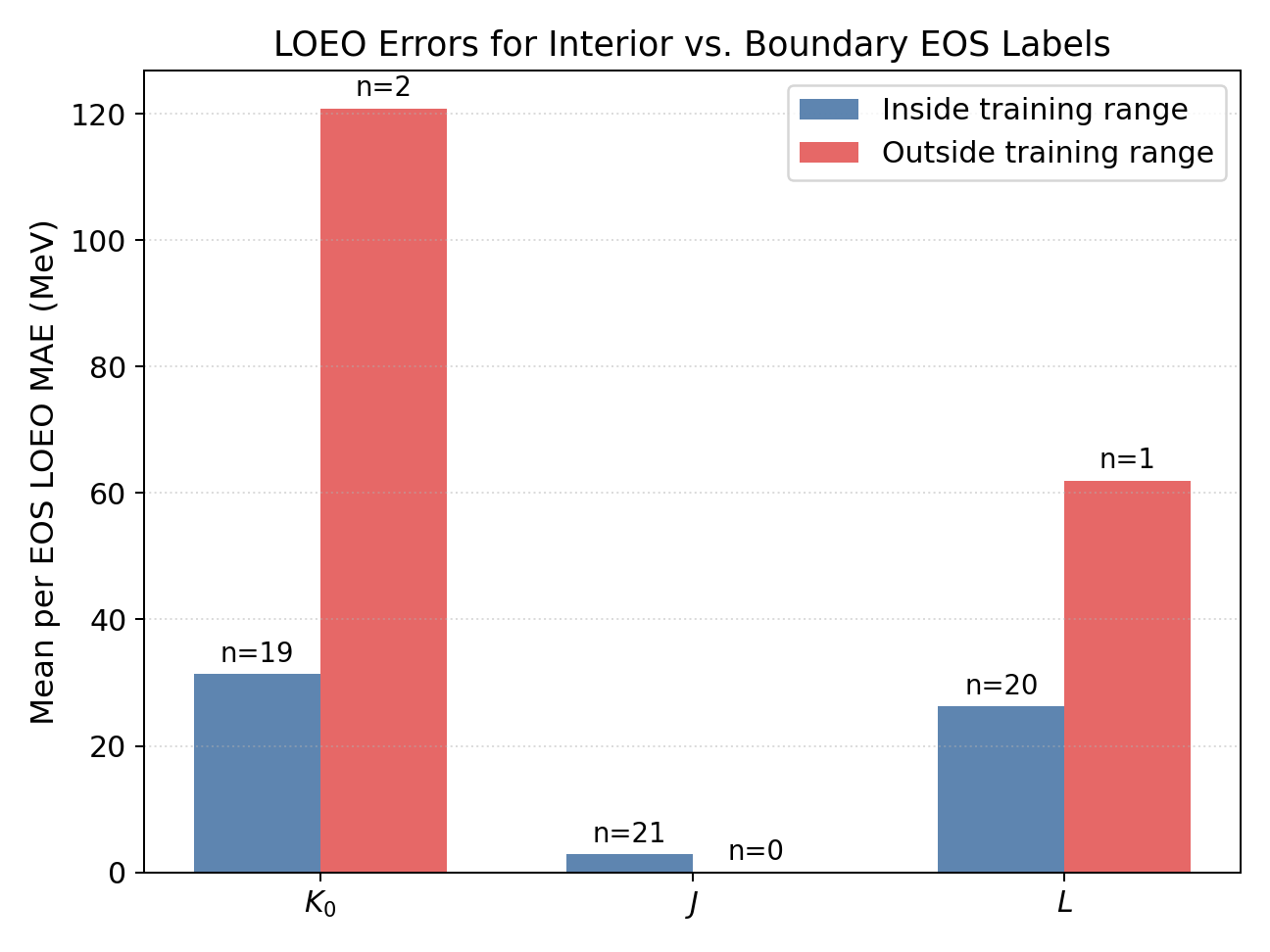}
    \caption{Mean per EoS LOEO MAE for raw time domain LightGBM predictions, split by whether the held out target value lies inside or outside the training target range. Boundary extrapolation is worst, but interior folds remain inaccurate. The outside training range bar for $J$ is zero because no held out $J$ target lies outside the training range.}
    \label{fig:loeo_regimes}
\end{figure}

However, the failure is not only a boundary effect. Figure~\ref{fig:loeo_regimes} separates held out EoS labels that lie inside the training target range from those outside it. For raw time domain LightGBM, interior EoS labels with no duplicate target triplet still have mean per EoS MAE $(47.47,3.28,16.98)$ MeV for $(K_0,J,L)$. Even labels whose $(K_0,J,L)$ triplet is duplicated in the training set are not always recovered: duplicated triplet folds give mean per EoS MAE $(27.10,2.71,29.34)$ MeV. HSDD2, for example, has $(K_0,J,L)=(243.0,31.7,55.0)$ MeV, a target triplet also present for BHBL and BHBLP, yet its waveform distribution does not transfer cleanly. The small catalogue therefore matters, but the diagnostic also indicates a genuine waveform family transfer problem.

\begin{figure}[t]
    \centering
    \includegraphics[width=\columnwidth]{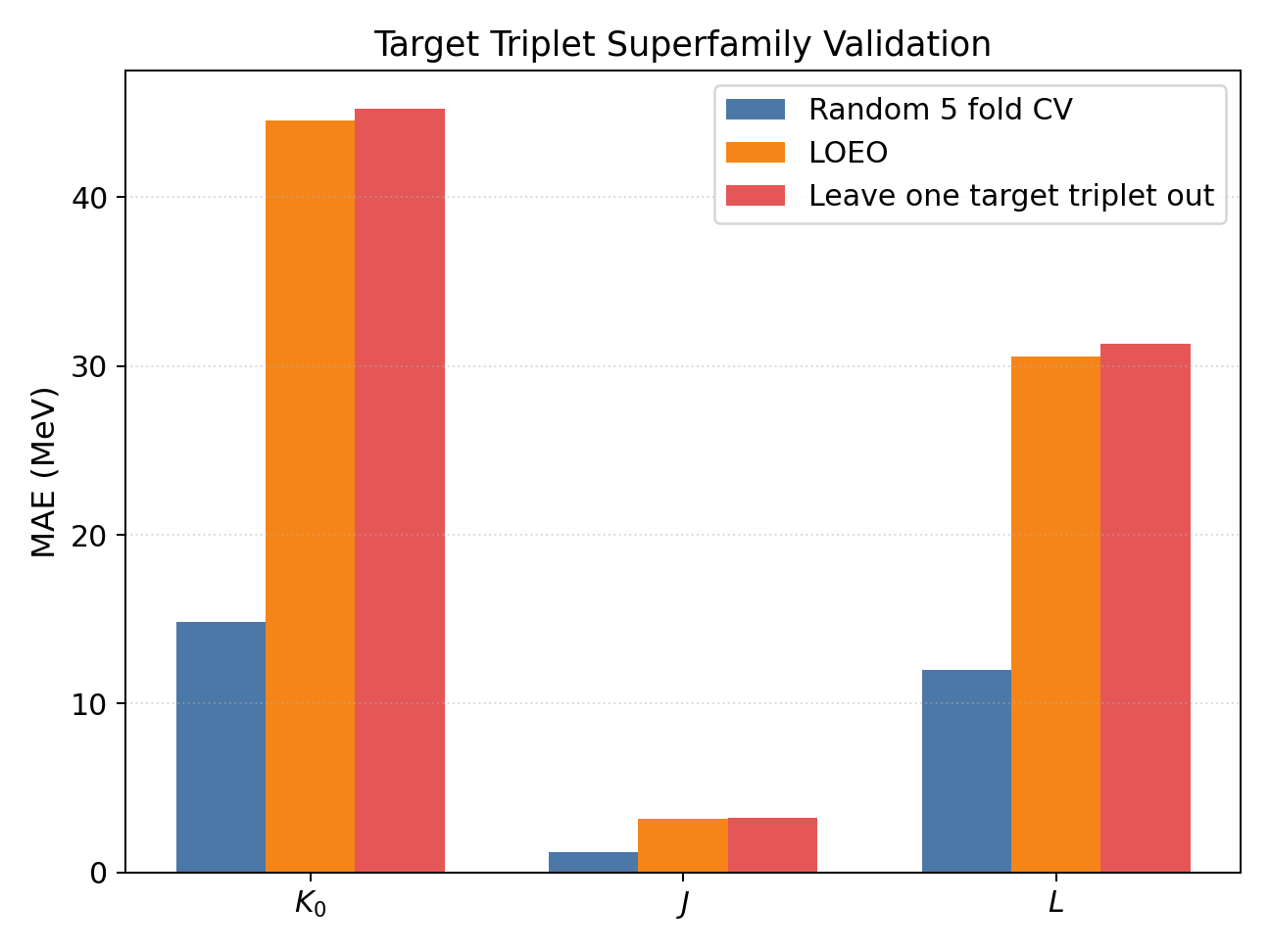}
    \caption{MAE comparison for raw time domain LightGBM under random CV, LOEO, and Leave One Target Triplet Out validation. The superfamily split holds out all EoS labels sharing the same $(K_0,J,L)$ target triplet.}
    \label{fig:triplet_superfamily}
\end{figure}

As a controlled check on the duplicated target issue, we also group all EoS labels sharing the same $(K_0,J,L)$ triplet into target triplet superfamilies and perform Leave One Target Triplet Out validation. This creates 11 held out superfamilies and ensures that no EoS label with the same saturation parameter triplet appears in the training set. The raw time domain MAEs are $(45.27,3.22,31.31)$ MeV for $(K_0,J,L)$, nearly identical to or worse than LOEO. Figure~\ref{fig:triplet_superfamily} shows that the performance loss remains when duplicated target values are removed, supporting the interpretation that sparse target coverage and waveform family transfer both contribute to the failure.

\section{Physical Feature Extraction}
To prevent the model from memorising high dimensional waveform geometries, we extract three low dimensional physical observables from the time series strain. These are not intended to be an exhaustive sufficient statistic for the CCSN signal. They are deliberately simple, physically interpretable diagnostics chosen to test whether the apparent random CV performance survives after removing most template shape degrees of freedom:
\begin{enumerate}
    \item \textbf{Bounce Amplitude ($A_{\text{bounce}}$)}: The peak negative strain amplitude, identified as the minimum strain value within the bounce window.
    \item \textbf{Bounce Width ($w_{\text{bounce}}$)}: The Full Width at Half Maximum (FWHM) of the main bounce peak, computed as the duration around $t_b$ where the strain is more negative than half of the bounce amplitude.
    \item \textbf{Peak g mode Frequency ($f_{\text{peak}}$)}: The peak frequency of post bounce oscillations in the $300\text{ Hz}$ to $2000\text{ Hz}$ band. This is extracted by taking the Fourier transform of the post bounce waveform (from $2$ to $6$ ms relative to $t_b$) and finding the frequency of peak power.
\end{enumerate}
All features are standardised using a standard scaler before model training. These hand extracted observables have unavoidable limitations: the FWHM can be noisy for weak or multi peaked bounce signals, and because the post bounce window used here is short, the extracted $f_{\text{peak}}$ should be interpreted as a compact early time spectral proxy rather than a precision measurement of the long time PNS oscillation track. We also compare it against the catalogue $f_{\rm peak}$ field when diagnosing correlations.

\begin{figure}[t]
    \centering
    \includegraphics[width=\columnwidth]{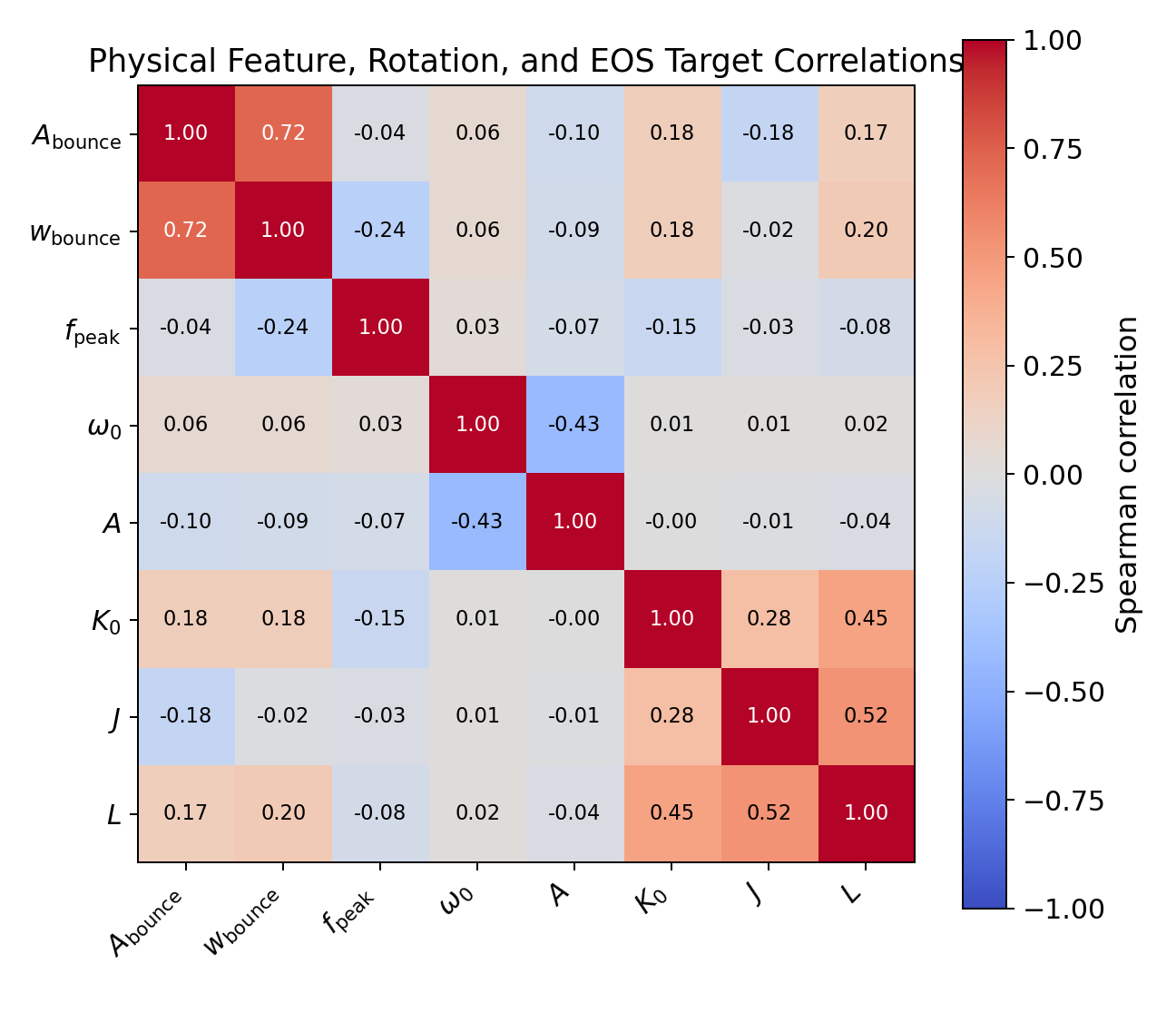}
    \caption{Spearman correlation matrix for extracted physical features, rotation variables, and EoS saturation targets. The direct monotonic correlations between the three extracted observables and $(K_0,J,L)$ are weak, consistent with a degenerate inverse problem.}
    \label{fig:feature_corr}
\end{figure}

\begin{table*}[t]
\caption{Regression $R^2$ scores under Random CV (with fold to fold standard deviation) vs. Leave One EoS Out (LOEO) CV, along with the corresponding LOEO Mean Absolute Error (MAE, in MeV), for multiple models using principal component analysis (PCA) frequency features vs. physical observables. Preprocessing (including target standardisation, feature scaling, and PCA projection to 5 components) is performed strictly inside each cross validation fold to prevent data leakage. Metrics are computed globally over all held out EoS predictions pooled together from each of the 21 folds.}
\label{tab:phys}
\begin{ruledtabular}
\begin{tabular}{llcccccc}
Target & Model & $R^2_{\text{Random}}$ (PCA) & $R^2_{\text{LOEO}}$ (PCA) & MAE (PCA) & $R^2_{\text{Random}}$ (Phys) & $R^2_{\text{LOEO}}$ (Phys) & MAE (Phys) \\
\midrule
$K_0$    & Ridge        & $0.1076 \pm 0.0351$ & -0.2572 & 34.09 & $0.1622 \pm 0.0622$ & -0.2828 & 34.19 \\
         & RandomForest & $0.3044 \pm 0.1051$ & -0.9576 & 42.03 & $0.4874 \pm 0.0783$ & -0.7831 & 40.97 \\
         & MLP          & $0.2001 \pm 0.0876$ & -0.7752 & 39.97 & $0.4031 \pm 0.1046$ & -0.5829 & 38.89 \\
         & LightGBM     & $0.2917 \pm 0.1103$ & -0.7741 & 40.00 & $0.4913 \pm 0.0613$ & -0.6353 & 38.43 \\
\midrule
$J$      & Ridge        & $0.1292 \pm 0.0263$ & -0.1162 &  2.67 & $0.0739 \pm 0.0314$ & -0.1715 &  2.79 \\
         & RandomForest & $0.2367 \pm 0.0418$ & -0.4658 &  3.03 & $0.3357 \pm 0.0598$ & -0.8852 &  3.45 \\
         & MLP          & $0.1920 \pm 0.0283$ & -0.2939 &  2.89 & $0.2047 \pm 0.0425$ & -0.7123 &  3.35 \\
         & LightGBM     & $0.2719 \pm 0.0313$ & -0.2906 &  2.87 & $0.3294 \pm 0.0131$ & -0.5562 &  3.18 \\
\midrule
$L$      & Ridge        & $0.0357 \pm 0.0232$ & -0.2538 & 26.25 & $0.0096 \pm 0.0146$ & -0.2183 & 26.33 \\
         & RandomForest & $0.0844 \pm 0.0357$ & -0.6425 & 29.81 & $0.1777 \pm 0.0614$ & -0.8694 & 31.53 \\
         & MLP          & $0.0530 \pm 0.0441$ & -0.5761 & 29.15 & $0.0952 \pm 0.0475$ & -0.6966 & 31.25 \\
         & LightGBM     & $0.1116 \pm 0.0230$ & -0.4875 & 28.65 & $0.2237 \pm 0.0274$ & -0.5425 & 29.35 \\
\end{tabular}
\end{ruledtabular}
\end{table*}

The performance of the models is summarised in Table~\ref{tab:phys}. Across all regressors (Ridge, RF, LightGBM, MLP), forcing the model to use the low dimensional physical observables reduces the Random CV performance (representing the removal of template leakage) and in several cases narrows the generalisation gap. However, the out of distribution LOEO $R^2$ remains negative in all configurations, and the LOEO MAE remains comparable to or worse than mean predictor baselines. 

The results suggest that this persistent failure is rooted in physical degeneracies, compounded by the sparse catalogue. As shown by Richers et al.~\cite{richers2017}, the bounce waveform is largely EoS independent and is controlled primarily by $T/|W|$ and differential rotation; clearer EoS dependence appears in the post bounce oscillation frequency, where $f_{\text{peak}}\sim 600$ to $1000$ Hz tracks the EoS dependent dynamical frequency of the core. Figure~\ref{fig:feature_corr} quantifies the problem in the present data. Because many waveforms share each EoS target, these correlations should be read as descriptive catalogue diagnostics rather than independent significance tests. The extracted $f_{\text{peak}}$ has weak Spearman correlations with $(K_0,J,L)$, $(-0.15,-0.03,-0.08)$, and even the catalogue $f_{\rm peak}$ field reaches only $(-0.15,-0.33,-0.33)$. Bounce amplitude correlates only weakly with $K_0$ (Spearman $0.18$, Pearson $0.34$) and weakly with $J$ and $L$. Thus the features identified by physical intuition and by the ML model are relevant, but they are not uniquely diagnostic of the EoS target once rotation and waveform family variation are allowed to change. This separates two failure modes: the present catalogue is too sparse for robust continuous regression, and the available GW observables are themselves degenerate without additional physical information.

We also consider a 21 class discrete EoS classification task on raw frequencies, following the classification setup evaluated by Abylkairov et al. (2024)~\cite{abylkairov2024}. Under Random 5 Fold CV, the classifier reaches an accuracy of $67.12\%$. However, under GroupKFold grouped by the differential rotation parameter $A$, where all samples from the same physical group are kept in the same fold, the classification accuracy collapses to $41.61\%$. This confirms that template leakage is not restricted to regression, but also dominates the apparent success of discrete classification pipelines.

\section{Explainable AI and SHAP Analysis}
\begin{figure}[t]
    \centering
    \includegraphics[width=\columnwidth]{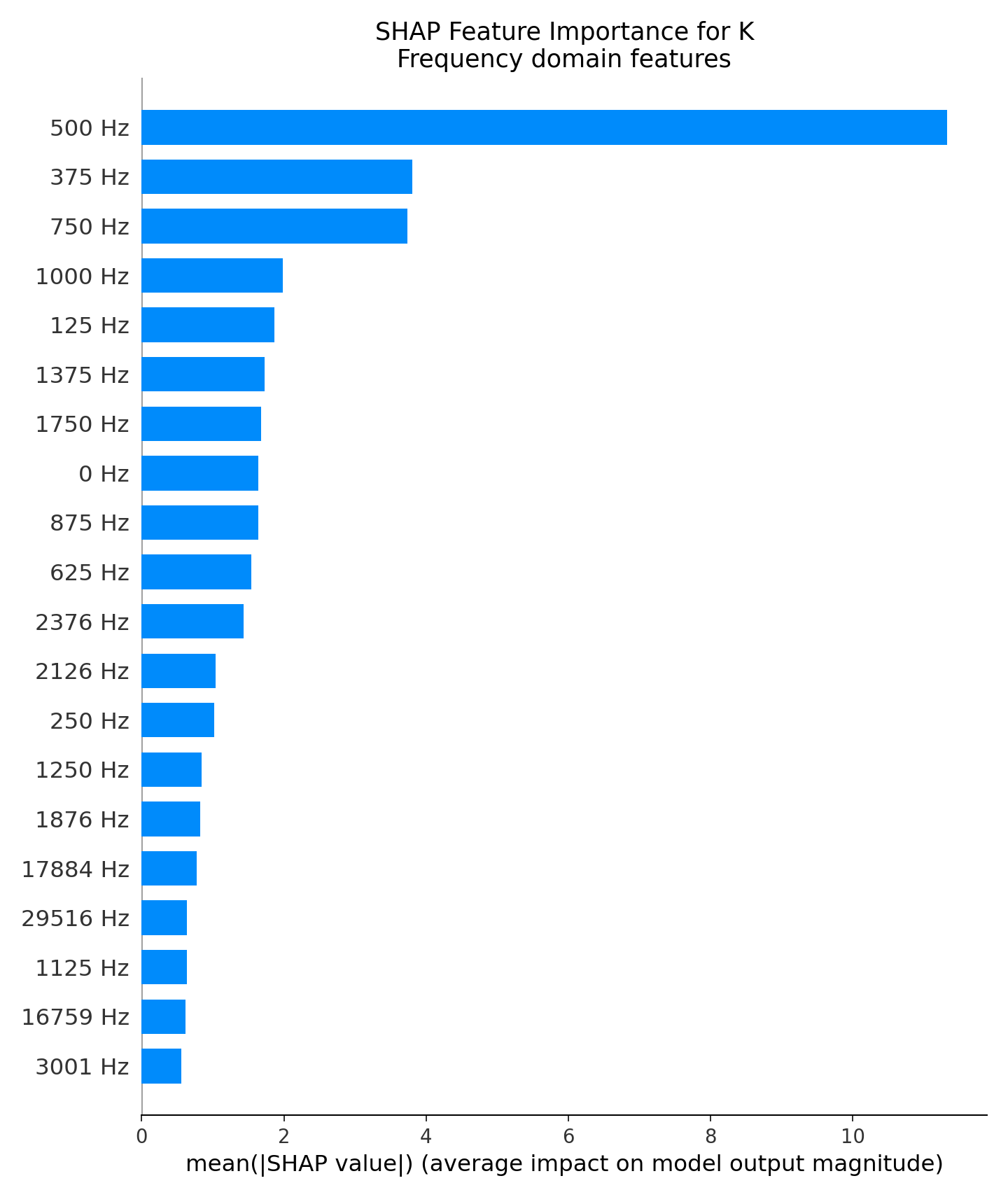}
    \caption{SHAP frequency importance for $K_0$ prediction. The model focuses heavily on features below $1$ kHz (representing the bounce envelope) and the $875$ to $1000\text{ Hz}$ band (PNS g modes).}
    \label{fig:shap}
\end{figure}

As a supporting interpretability check, we compute SHAP (SHapley Additive exPlanations) values~\cite{lundberg2017} for the frequency domain LightGBM model and map fast Fourier transform (FFT) bins back to physical Hertz using $\Delta f=f_s/524\simeq125$ Hz, where $f_s=65.5\text{ kHz}$ is the waveform sampling frequency and 524 is the number of time samples in the cropped bounce window. SHAP assigns an importance value to each input feature for a model prediction, allowing us to ask which frequency bins the model relies on. As shown in Fig.~\ref{fig:shap}, the most important bins lie from $375$ to $500$ Hz, associated with the bounce envelope, and from $875$ to $1000$ Hz, associated with early PNS g mode oscillations. This is reassuring in the limited sense that the model uses physically relevant frequency bands. It does not imply a transferable EoS estimator: those same bands are controlled by rotation, bounce compactness, and catalogue specific waveform family structure. SHAP therefore supports the main interpretation that random CV success arises from physically meaningful but non unique observables, not from a robust inverse EoS map.

\section{Progenitor Mass Classification}
We contrast the EoS regression with Progenitor Mass Classification ($12, 15, 27, 40 M_\odot$) on the Mitra et al. (2022) catalogue~\cite{mitra2022}. The dataset contains 402 waveforms in total, with a balanced class distribution: $12 M_\odot$ (97 samples), $15 M_\odot$ (99 samples), $27 M_\odot$ (102 samples), and $40 M_\odot$ (104 samples). The waveforms span 32 unique rotation speeds ($\omega_0$ from \texttt{O01} to \texttt{O16.5}) and 5 unique differential rotation profiles ($A$ from \texttt{A1} to \texttt{A5}). We evaluate how well the mass prediction generalises to entirely unseen rotation configurations using GroupKFold grouped by angular velocity $\omega_0$ and differential rotation profile $A$.

\begin{figure}[t]
    \centering
    \includegraphics[width=\columnwidth]{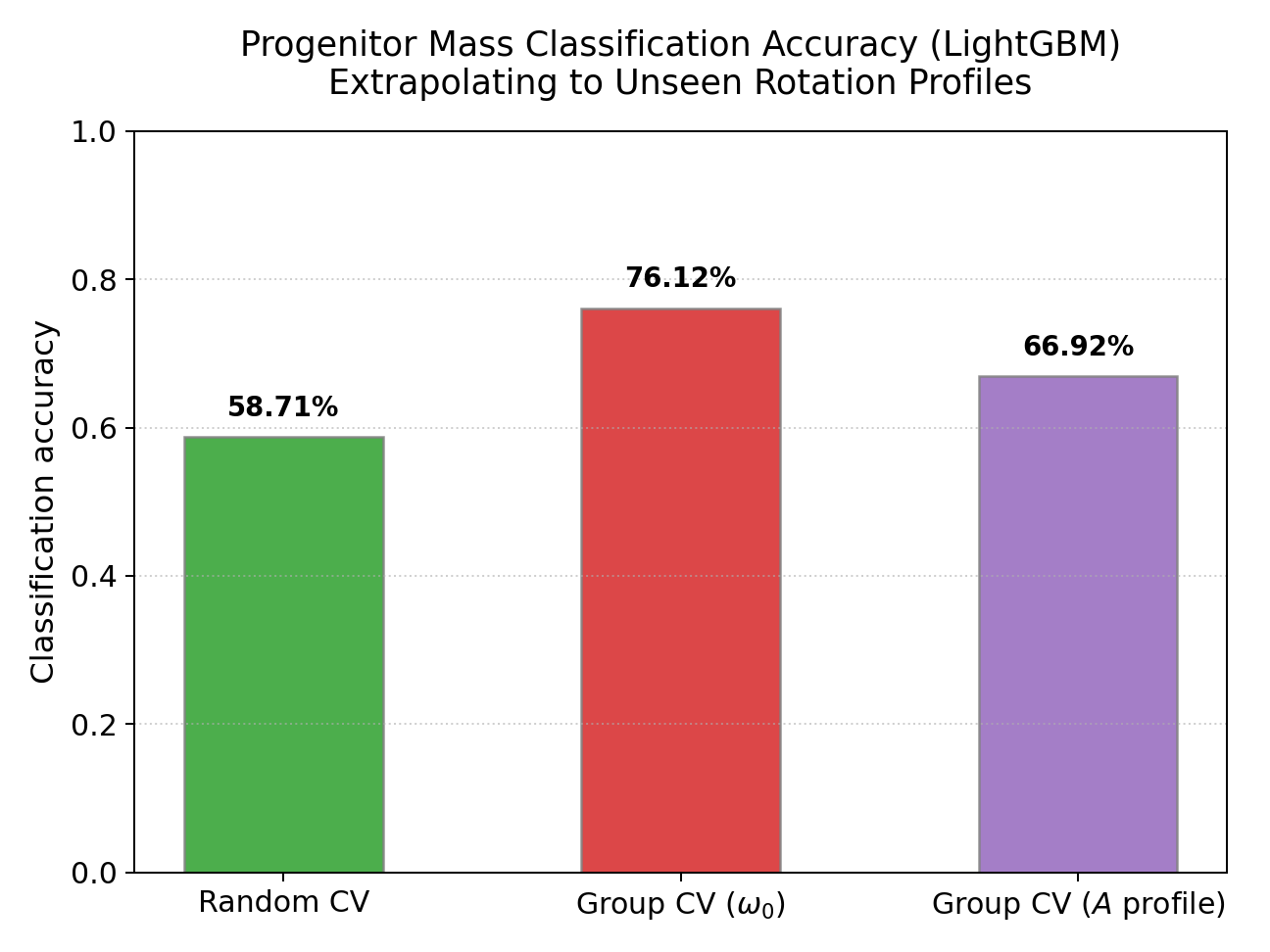}
    \caption{LightGBM accuracy for progenitor mass classification across Random CV and Group CV (unseen rotation rate $\omega$ and profile $A$ splits).}
    \label{fig:progenitor}
\end{figure}

As shown in Fig.~\ref{fig:progenitor}, the mass classifier generalises exceptionally well to unseen rotation profiles:
\begin{itemize}
    \item \textbf{Random 5 Fold CV Accuracy}: $58.71\%$
    \item \textbf{GroupKFold (grouped by $\omega_0$)}: $\mathbf{76.12\%}$
    \item \textbf{GroupKFold (grouped by A profile)}: $\mathbf{66.92\%}$
\end{itemize}

\begin{figure*}[t]
    \centering
    \includegraphics[width=0.95\textwidth]{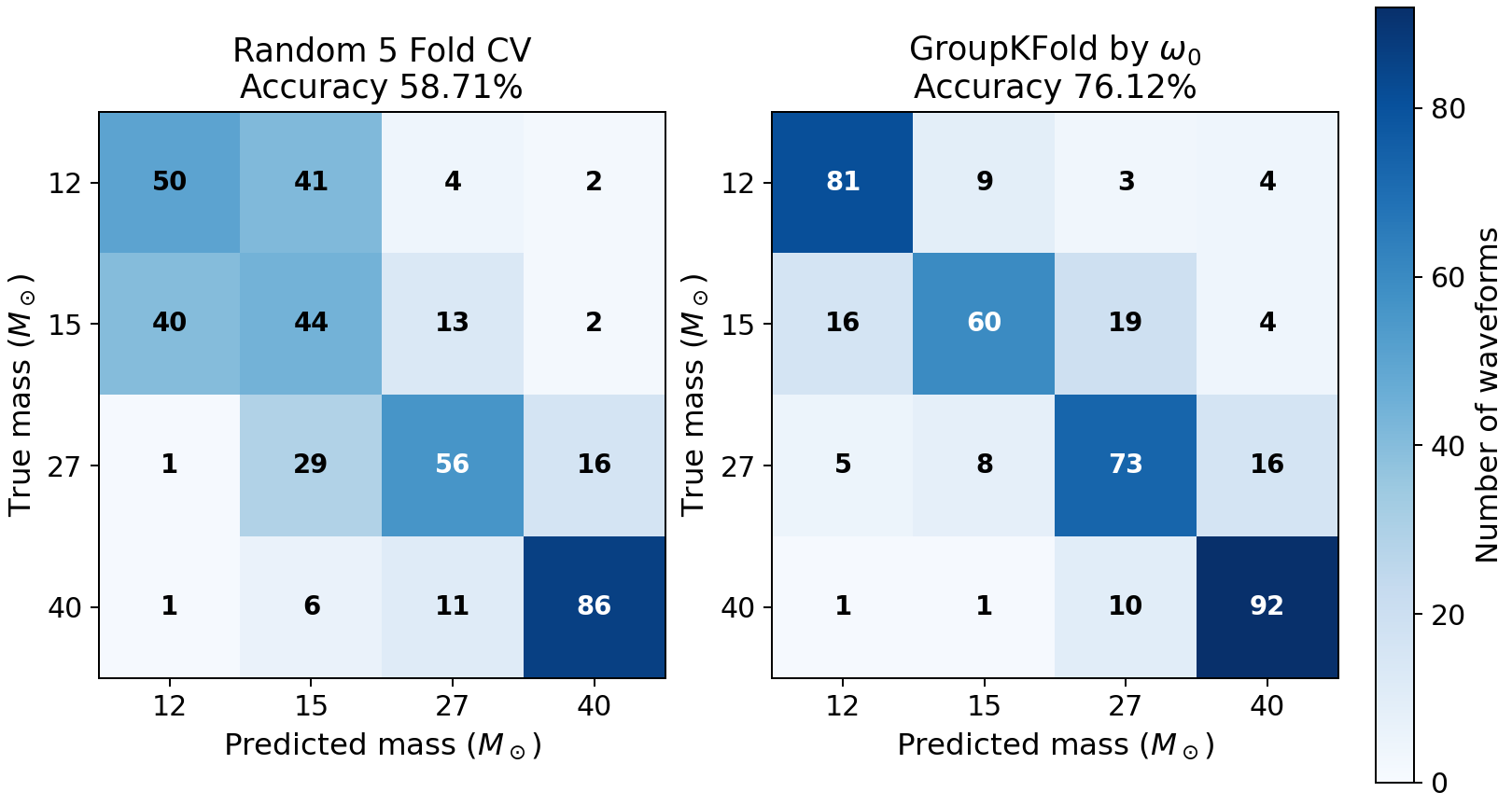}
    \caption{Confusion matrices for progenitor mass classification under Random 5 Fold CV and GroupKFold by $\omega_0$. Rows represent the true mass and columns represent the predicted mass. The stronger diagonal in the grouped split shows that holding out rotation speeds reduces adjacent mass confusion.}
    \label{fig:confusion}
\end{figure*}

Unlike EoS regression, progenitor mass classification has no generalisation gap under unseen rotation splits; in fact, grouping by rotation speed improves performance by $17.4\%$. Under Random CV, the training and testing sets share the same rotation values ($\omega_0$ and $A$). The model attempts to classify progenitor mass, but is confounded by the shared rotation parameters. This is shown in Fig.~\ref{fig:confusion}: under Random CV, there is substantial confusion between adjacent masses (specifically $12$ vs $15\ M_\odot$, and $15$ vs $27\ M_\odot$), because similar rotation rates produce similar frequency modulations.

When the split is grouped by $\omega_0$, the test fold contains entirely unseen rotation velocities. The model is forced to ignore the rotation specific fine structures. Instead, it relies on global envelope features, such as the overall amplitude and duration of the bounce, which scale with the mass and compactness of the progenitor star. These global features are robust and generalise well, leading to a much higher classification accuracy ($76.12\%$). This suggests that progenitor mass controls the macrodynamics, creating distinct wave envelopes that generalise well. In contrast, EoS parameters control the microphysics, producing subtle modifications that are easily degenerate under varying differential rotation profiles.

\subsection{Progenitor Mass Regression and Out of Sample Mass Inversion}
While mass classification generalises well across unseen rotation speeds of the same catalogue progenitors, a key open question is whether machine learning models can estimate the mass of an entirely \textit{unseen} progenitor mass. To address this, we reformulate mass inversion as a continuous regression problem ($M \in \{12, 15, 27, 40\}\ M_\odot$) on the multi progenitor catalogue and evaluate Ridge Regression and LightGBM under three validation protocols:
\begin{enumerate}
    \item \textbf{Random 5 Fold CV}: Shuffling all waveforms randomly.
    \item \textbf{GroupKFold (by $\omega_0$)}: Evaluating generalisability to unseen rotation speeds.
    \item \textbf{Leave One Mass Out (LOMO) CV}: The model is trained on three progenitor masses and tested on the fourth remaining mass. This is repeated for all four masses to compute the overall out of distribution performance.
\end{enumerate}

The regression metrics are summarised in Table~\ref{tab:mass_regression}. Under Random 5 Fold CV, LightGBM achieves a high performance of $R^2 = 0.7226$ (MAE = $4.28 M_\odot$). This accuracy remains robust under GroupKFold by $\omega_0$ ($R^2 = 0.6975$, MAE = $4.45 M_\odot$), demonstrating that the models can interpolate across rotation configurations when the progenitor mass is inside the training distribution. 

\begin{figure}[t]
    \centering
    \includegraphics[width=\columnwidth]{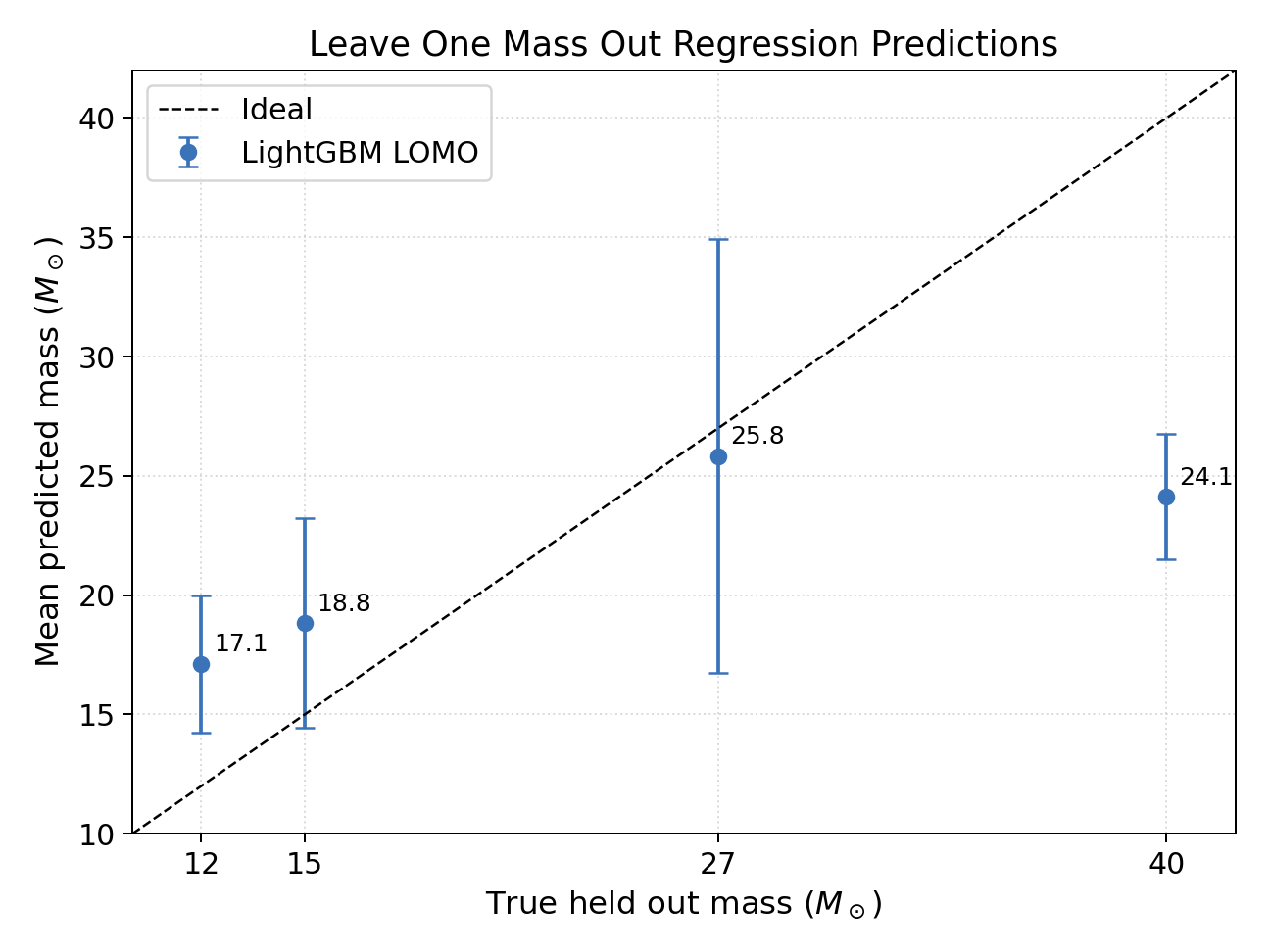}
    \caption{Mean LightGBM predictions under Leave One Mass Out regression. Error bars show the standard deviation across waveforms of the held out mass. The endpoint compression, especially for the $40 M_\odot$ model, shows that the model interpolates within the sparse mass grid rather than performing reliable out of distribution inversion.}
    \label{fig:mass_lomo}
\end{figure}

However, under LOMO CV, the regression performance degrades substantially ($R^2 = 0.1498$, MAE = $8.50 M_\odot$ for LightGBM; $R^2 = -0.0815$, MAE = $9.31 M_\odot$ for Ridge Regression). With only four mass values, pooled $R^2$ is again only a compact diagnostic. The held out predictions in Fig.~\ref{fig:mass_lomo} show the more physical failure mode: LightGBM predicts mean masses of $17.13$, $18.83$, $25.83$, and $24.12 M_\odot$ when the true held out masses are $12$, $15$, $27$, and $40 M_\odot$, respectively. The model overpredicts the low mass endpoints and severely underpredicts the $40 M_\odot$ endpoint, indicating regression towards the interior of the sparse mass grid rather than reliable continuous mass inversion.

\begin{table}[t]
\caption{Progenitor mass regression metrics ($R^2$ and MAE in $M_\odot$) across Random CV, GroupKFold by $\omega_0$, and Leave One Mass Out (LOMO) CV on the multi progenitor catalogue.}
\label{tab:mass_regression}
\begin{ruledtabular}
\begin{tabular}{llccc}
Model & Metric & Random CV & Group $\omega_0$ & LOMO CV \\
\midrule
Ridge & $R^2$ & 0.3034 & 0.3518 & -0.0815 \\
      & MAE & 6.9098 & 6.8234 & 9.3098 \\
\midrule
LightGBM & $R^2$ & 0.7226 & 0.6975 & 0.1498 \\
         & MAE & 4.2831 & 4.4480 & 8.4975 \\
\end{tabular}
\end{ruledtabular}
\end{table}

\section{Implications for Third Generation Detector Pipelines}
Future third generation (3G) ground based gravitational wave observatories, such as the Einstein Telescope (ET) and Cosmic Explorer (CE), will drastically improve the prospects of detecting and characterising CCSNe. The literature values summarised in Table~\ref{tab:detector_context} show why this is a high value regime for ML assisted inference: 3G detectors may measure rotation and PNS oscillation frequencies with useful precision for Galactic or Magellanic Cloud events, and may extend EoS classification horizons beyond the Milky Way for optimally oriented rapidly rotating sources. A high SNR, however, does not by itself resolve the out of distribution generalisation failure identified in this work.

\begin{table*}[t]
    \caption{Representative detector context values from the cited literature. These values motivate the importance of CCSN GW inference for 3G detectors, but they do not remove the need for out of distribution validation.}
\label{tab:detector_context}
\begin{ruledtabular}
\begin{tabular}{p{0.23\textwidth}p{0.31\textwidth}p{0.36\textwidth}}
Study & Published value & Relevance to this work \\
\midrule
Srivastava et al.~\cite{srivastava2019} & A supernova optimised CE like detector remains limited to roughly $\mathcal{O}(100)$ kpc for likely CCSN waveforms; a one per year CCSN GW rate would require strain sensitivity near $3\times 10^{-27}\,{\rm Hz}^{-1/2}$ over $100$ to $1500$ Hz. & The relevant detector band overlaps the $f_{\rm peak}$ range used by EoS classifiers and regressors, but event rarity means validation errors are costly. \\
Afle and Brown~\cite{afle2020} & For a rapidly rotating source at $8.1$ kpc, CE can constrain $\beta$ to about $8\times10^{-4}$ and post bounce frequency to about $5$ Hz at 90\% credibility; at $48.5$ kpc these degrade to about $0.003$ and $11$ Hz. & 3G detectors may measure the physical observables used in this paper precisely enough for inference, but the mapping from observables to EoS remains degenerate. \\
Abylkairov et al.~\cite{abylkairov2025} & For optimally oriented sources, EoS classification can remain above $\sim70\%$ to about $20$ kpc for A+, $80$ kpc for ET, and $100$ kpc for CE; for random orientations, ET/CE performance drops below $\sim70\%$ beyond about $30$ kpc. & Distance and orientation effects are coupled to the same waveform features that drive apparent catalogue performance. \\
\end{tabular}
\end{ruledtabular}
\end{table*}

\subsection{Detector Noise Stress Test}
\begin{figure}[t]
    \centering
    \includegraphics[width=\columnwidth]{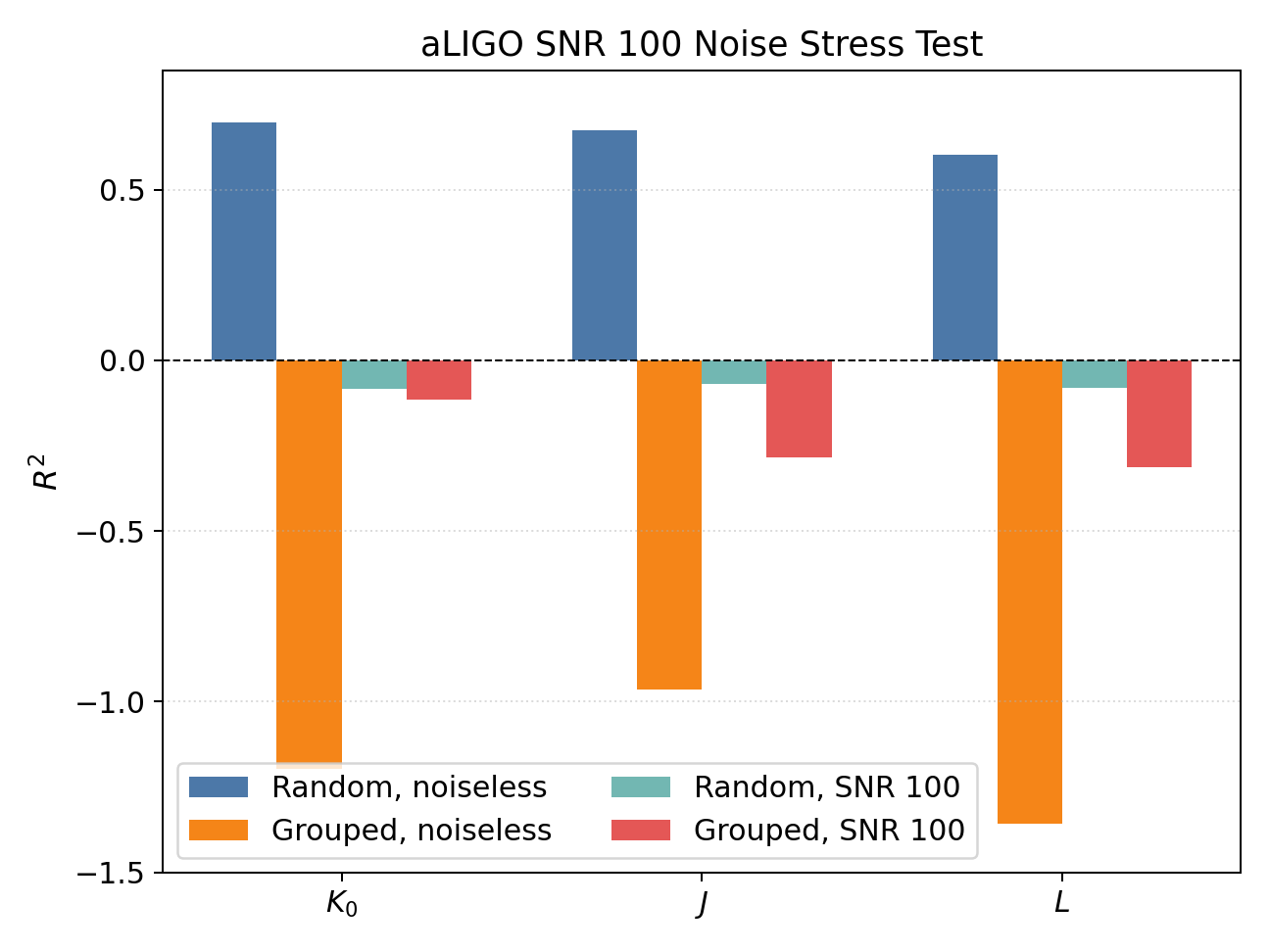}
    \caption{Time domain LightGBM $R^2$ scores under noiseless and aLIGO SNR 100 conditions. Detector noise removes the apparent random CV success and keeps grouped validation below zero.}
    \label{fig:noise_stress}
\end{figure}

To evaluate the effect of detector noise, we repeat our evaluation under Advanced LIGO (aLIGO) Zero Detuning High Power noise injection, using the design power spectral density curve for this configuration~\cite{ligoT0900288}, at a matched filter signal to noise ratio (SNR) of 100. The quantitative metrics are shown in Table~\ref{tab:noise} and Fig.~\ref{fig:noise_stress}. The striking result is that noise suppresses even the random CV performance: raw time domain random CV scores become $R^2=(-0.083,-0.069,-0.079)$ for $(K_0,J,L)$, while grouped scores remain negative. This suggests that the template level signal exploited in noiseless random CV is fragile under detector perturbations. The clean LOEO tests show that out of distribution generalisation is already poor before realistic detector complications are included; the noisy tests show that apparent catalogue interpolation can also be fragile once the waveform is perturbed.

\begin{table}[t]
\caption{LightGBM regression $R^2$ scores under noiseless vs. noisy (aLIGO, matched filter SNR = 100) conditions for raw time domain and frequency domain representations, comparing Random CV vs. Group CV (grouped by EoS).}
\label{tab:noise}
\begin{ruledtabular}
\begin{tabular}{lllcc}
Target & Representation & Noise Case & $R^2_{\text{Random}}$ & $R^2_{\text{Group}}$ \\
\midrule
$K_0$ & Time Domain & Noiseless & 0.698 & -1.197 \\
      &             & aLIGO SNR 100 & -0.083 & -0.114 \\
      & Freq Domain & Noiseless & 0.656 & -0.994 \\
      &             & aLIGO SNR 100 & -0.086 & -0.150 \\
\midrule
$J$   & Time Domain & Noiseless & 0.674 & -0.964 \\
      &             & aLIGO SNR 100 & -0.069 & -0.283 \\
      & Freq Domain & Noiseless & 0.555 & -0.910 \\
      &             & aLIGO SNR 100 & -0.071 & -0.272 \\
\midrule
$L$   & Time Domain & Noiseless & 0.601 & -1.357 \\
      &             & aLIGO SNR 100 & -0.079 & -0.314 \\
      & Freq Domain & Noiseless & 0.468 & -1.316 \\
      &             & aLIGO SNR 100 & -0.081 & -0.337 \\
\end{tabular}
\end{ruledtabular}
\end{table}

We note that the present work uses the aLIGO noise curve as a preliminary stress test to establish the baseline of noise impact; ET and CE specific noise injections and reconstruction pipelines are left for future work. A realistic 3G analysis should repeat the grouped validation audit using Einstein Telescope D and CE power spectral densities (PSDs), uncertain source distance and orientation, imperfect bounce time recovery, and waveform reconstruction uncertainties from burst pipelines. If an ML pipeline is trained on a finite simulation catalogue and deployed on real 3G data, it may produce overconfident dense matter parameter estimates unless it has been audited under out of distribution conditions. LOEO and grouped validation by physical template families should therefore be treated as standard reporting requirements for ML based CCSN parameter estimation pipelines.

The practical path forward is not to abandon ML, but to change what is being learned and how catalogues are built. Promising directions include simulation based inference with explicit priors over nuisance parameters, physics informed architectures that condition on rotation and progenitor information rather than silently absorbing them, active learning to add simulations where LOEO residuals are largest, and multi messenger conditioning using electromagnetic progenitor constraints or neutrino informed rotation estimates. In particular, neural posterior, likelihood, or likelihood ratio estimators could be trained on simulation catalogues while marginalising over rotation, progenitor structure, detector response, and noise realisations; their calibration should then be tested with the same leave family out protocols used here. In this framing, ML becomes an emulator or likelihood component inside a physics aware inference pipeline, rather than a direct waveform to EoS lookup table.

\subsection{Methodological Limitations and Future Work}
We note several key limitations of the current study:
\begin{enumerate}
    \item \textbf{Catalogue Resolution}: The simulation catalogue used for EoS regression contains 21 EoS labels but only 11 unique $(K_0,J,L)$ target triplets, and the multi progenitor catalogue contains only 4 progenitor masses. These are standard benchmark sets in the ML literature, but they are too sparse to infer how many EoS models would be required for positive LOEO performance. Testing on higher resolution, continuous grids of dense matter parameters, including modern 3D calculations that explicitly compare EoS dependent explosion and GW signatures~\cite{rusakov2026}, would provide a more detailed map of the generalisation boundaries.
    \item \textbf{Simulation Code and Microphysics Systematics}: The Richers waveforms are consistent with the CoCoNuT simulation lineage used in the Dimmelmeier and Abdikamalov rotating collapse studies~\cite{dimmelmeier2002,dimmelmeier2005,abdikamalov2014,richers2017}. The important caveat is what that lineage assumes: CFC treats the spatial metric as conformally flat, so freely propagating GW degrees of freedom are not evolved directly and the strain is reconstructed with a quadrupole formula. The simulations also use a $Y_e(\rho)$ prescription for electron capture during collapse and approximate leakage after bounce. These approximations are appropriate for the early bounce/ringdown audit performed here, but they are additional systematics that should be cited when interpreting ML performance as dense EoS inference.
    \item \textbf{Dimensionality and Symmetry}: The waveforms analysed here are generated from 2D axisymmetric general relativistic hydrodynamic simulations. Extensions to full 3D simulations (which introduce stochastic neutrino driven convection, viewing angle dependence, memory components, and non axisymmetric modes) will be essential to verify if the generalisation failure is further compounded by stochastic signal components. A natural follow up is to repeat the same leave family out audit on the public long term 3D Burrows/Princeton waveform set, including matter strains, quadrupole tensors, neutrino memory, and angular dependence~\cite{vartanyan2023,choi2024,burrowsGWData}.
    \item \textbf{Detector Noise Simplification}: The noise injection evaluated here serves as a preliminary baseline stress test using aLIGO noise. Real world deployment on Einstein Telescope (ET) or Cosmic Explorer (CE) pipelines will require repeating the evaluation under 3G detector noise, signal reconstruction uncertainties (e.g., using coherent WaveBurst or BayesWave), and sky localisation variations.
\end{enumerate}
Under out of distribution validation, the systematic failure found here indicates that the available waveform features are insufficient to learn a transferable mapping within the present catalogue. This does not prove that GW only EoS inference is fundamentally impossible, but it does show that present benchmark catalogues and standard random CV protocols are not enough to establish it.

\section{Discussion and Conclusions}
These physical degeneracies are consistent with traditional parameter estimation methods. For example, Pastor Marcos et al. (2024)~\cite{pastormarcos2024} utilised Bayesian inference on fast rotating CCSN waveforms and noted that parameter extraction is heavily limited by intrinsic degeneracies in the waveform morphology, particularly between the nuclear EoS and the core rotation rate. Our machine learning audit shows that ML algorithms do not automatically bypass these physical limitations; under random CV they can interpolate within the catalogue by exploiting shared rotation profiles and template structure. Furthermore, our findings complement recent efforts to characterise CCSN signals through alternative formulations, such as gravitational wave entropy (Sakan et al. 2025~\cite{sakan2025}), which extracts Shannon, Rényi, Tsallis, and exponential entropy measures in the wavelet domain to compress waveforms and perform classification. These entropy measures may be useful as low dimensional summaries complementary to bounce amplitude, width, and post bounce frequency, especially for noisy burst reconstruction outputs. However, like learned waveform embeddings, they still require out of distribution validation before being interpreted as physical inference tools. Together with advanced detector reconstruction studies (Szczepa\'nczyk et al. 2021~\cite{szczepanczyk2021}, Gossan et al. 2016~\cite{gossan2016}), these results indicate that external astrophysical priors may be necessary before GW only regression of dense matter saturation constants becomes reliable.

In this work, we present a rigorous evaluation of machine learning models for CCSN GW parameter estimation:
\begin{enumerate}
    \item Standard random cross validation on raw waveforms yields misleadingly high performance due to template leakage. Future studies should report Leave One EoS Out or leave family out validation, with MAE and mean predictor baselines, to verify physical generalisability.
    \item Restricting models to low dimensional physical observables ($A_{\text{bounce}}$, $w_{\text{bounce}}$, $f_{\text{peak}}$) reduces apparent leakage and sometimes narrows the generalisation gap, but does not restore positive out of distribution performance. The feature target correlations remain weak, showing that the relevant observables are not uniquely diagnostic of EoS parameters in this catalogue.
    \item Progenitor mass can be classified robustly across unseen rotation speeds of the same progenitors (up to $76.12\%$ accuracy). However, continuous mass inversion evaluated with Leave One Mass Out validation compresses endpoint masses towards the catalogue interior, confirming that catalogue interpolation is much easier than out of distribution inference.
\end{enumerate}

To break the degeneracies of continuous EoS parameter estimation, future pipelines should incorporate multi messenger constraints, physics informed model structure, active catalogue expansion, simulation based inference, and validation protocols that hold out entire physical template families. The result of this paper is therefore a caution, not a pessimistic endpoint: ML may still be valuable for 3G CCSN science, but only if its validation matches the physical inference problem.

\begin{acknowledgments}
A.M. thanks Professor Adam Burrows for useful discussions.
\end{acknowledgments}

\end{document}